\documentclass[cits]{PoS}
\usepackage{epsfig}
\usepackage{amsmath}   

\newcommand{\bea}{\begin{eqnarray}}
\newcommand{\eea}{\end{eqnarray}}
\newcommand{\bqa}{\begin{eqnarray}}
\newcommand{\eqa}{\end{eqnarray}}

\newcommand{\nl}{\nonumber \\}

\newcommand{\eps}{\varepsilon}
\newcommand{\bq}{ \begin {equation} }
\newcommand{\eq}{\end{equation}}
\newcommand{\be}{\begin{eqnarray}}
\newcommand{\ea}{\end{eqnarray}}

\def \EulerGamma {\gamma_E}

\title{%
{\small DESY 08--007 \newline SFB/CPP--08--08 \newline HEPTOOLS 08--013} \\[1.5cm]
A new treatment of mixed virtual and real IR-singularities 
}

\ShortTitle{Mixed virtual and real IR-singularities}

\author{Janusz Gluza \\
        Department of Field Theory and Particle Physics,
    Institute of Physics, \\
    University of Silesia, Uniwersytecka 4, PL-40-007 Katowice,
    Poland
\\
       E-mail: \email{gluza@us.edu.pl}}

\author{\speaker{Tord Riemann} 
\\
   Deutsches Elektronen-Synchrotron, DESY,
   Platanenallee 6, 15738 Zeuthen, Germany
\\
   E-mail: \email{Tord.Riemann@desy.de}}

\abstract{We discuss the determination of the infrared singularities of massive one-loop 5-point functions with Mellin-Barnes (MB) representations.
Massless internal lines may lead to poles in the $\eps$ expansion of the Feynman diagram, 
while unresolved massless final state particles give endpoint singularities of the phase space integrals.
MB integrals are an elegant tool for their common treatment.
An evaluation by taking residues leads to inverse binomial sums.  
 }

\FullConference{8th International Symposium on Radiative Corrections (RADCOR)\\
		 October 1-5 2007\\
		 Florence, Italy}

\begin{document}
\allowdisplaybreaks

\section{Introduction}
We study the infrared (IR) singularities of some massive one-loop $n$-point Feynman integrals, 
\bea \label{eq-bha}
I = 
\frac{e^{\eps \gamma_E}}{i\pi^{d/2}} \int \frac{d^dk~~ T(k)}
     {(q_1^2-m_1^2)^{\nu_1} \ldots (q_i^2-m_i^2)^{\nu_j} \ldots
       (q_n^2-m_n^2)^{\nu_n}  }  ,
\eea
by representing them with standard Feynman parameter integrals with characteristic $F$ and $U$ forms:
\bea
\label{eq-scalar1}
I &=& 
\frac{e^{\eps \gamma_E}(-1)^{N_{\nu}} \Gamma\left(N_{\nu}-\frac{d}{2}\right)}
{\prod_{i=1}^{n}\Gamma(\nu_i)}
\int_0^1 \prod_{j=1}^n dx_j ~ x_j^{\nu_j-1}
\delta(1-\sum_{i=1}^n x_i)
\frac{U(x)^{N_{\nu}-d}}{F(x)^{N_{\nu}-d/2}}~P(T),
\nl
\eea
with
$N_{\nu} = \sum_{i=1}^{n} \nu_i$.
For one loop integrals, the $U=\sum x_i$ may be set to one.
The $F$-form is bilinear in the $x_i$ and may be represented in turn by a multiple Mellin-Barnes (MB) integral, using the representation:
\be
\frac{1}{(A(x)+B(x))^{\nu}}
= \frac{1}{2 \pi i}
\int\limits_{-i \infty+R}^{i \infty+R}
dz ~{A(x)^z ~
B(x)^{-\nu-z}} ~ \frac{\Gamma{(-z)}\Gamma{(\nu+z)} }{\Gamma{(\nu)}},
\label{mb}
\ea
where the integration contour separates the poles of the $\Gamma$-functions.

Afterwards, the Feynman parameters may be integrated out and one has to solve the resulting MB integral.
This is in general quite non-trivial.
However, there is an interesting kind of problems where a systematic approach might be developed, namely the evaluation of the IR divergent parts of the Feynman integrals.
They are at the begin of the $\eps$-expansion ($\eps = (4-d)/2$) of the Feynman integral and so of smaller dimensionality in the variables $z$. 
In fact,  usually one subtracts them from the rest of the integral and treats them separately.

The MB representation allows to do this in a special way which might be of some practical usefulness.
We will discuss here only scalar one-loop functions, $T(k)=1$, but  tensors don't show additional problems.
For basic definitions and formulae we refer to \cite{Gluza:2007rt,Gluza:2007bd,Gluza:2007uw} and references cited therein.
We use here and in the following the Mathematica packages AMBRE \cite{Gluza:2007rt} and MB \cite{Czakon:2005rk} for the derivations of the MB representations and for the $\eps$-expansions.
In section 2 we apply the MB-approach to the massive Bhabha vertex and box functions and extract their $\eps$-poles.
Section 3 contains the treatment of both the virtual $\eps$-poles and the endpoint singularities from an unresolved, massless particles in a pentagon diagram of  massive Bhabha scattering.
The method may be generalized to more complex cases, including higher loop orders, but explicit evaluations become then more and more complicated.
\section{Simple $\eps$-poles: Massive QED vertex and box}
We will set $m=1$, and $s$, $t$ are the usual Mandelstam variables.
The QED vertex function has the $F$-form:     
\bea
F(s) = [X[2] + X[3]]^2 + [- s]~X[2]X[3],
\eea
leading, without a continuation in $\eps$, to a one-dimensional MB representation and a series over residues
\cite{Gluza:2007bd}:
\bea\label{defv3la}
V(s)  &=& 
  \frac{1}{2s\eps} \frac{e^{\eps\gamma_E}}{2\pi i}\int\limits_{-i \infty-1/2}^{-i \infty-1/2} dz~~ (-s)^{-z}\frac{\Gamma^2(-z)\Gamma(-z+\eps)\Gamma(1+z)}
{\Gamma(-2z)}
\nl
&=&  - \frac{e^{\eps\gamma_E}} {2\eps} \sum_{n=0}^{\infty} \frac{s^n}{\binom{2n}{n}(2n+1)} 
\frac{\Gamma(n+1+\eps)}{\Gamma(n+1)}.
\eea
The complete series may be summed directly with Mathematica\footnote{The expression for $V(s)$ was also derived in \cite{Huber:2007dx}; see additionally \cite{Davydychev:2000na}.}, and the vertex becomes:
\bea
V(s) = -\frac {e^{\eps\gamma_E}} {2\eps} \Gamma(1+\eps) ~~_2F_1\left[1,1+\eps;3/2;s/4 \right] .
\eea
Alternatively, one may derive the $\eps$-expansion by exploiting the well-known relation with harmonic numbers $S_k(n) = \sum_{i=1}^{n}1/i^k$:
\bea
\label{harmsum}
\frac{\Gamma(n+a\eps)}{\Gamma(n)} &=& \Gamma(1+a\eps) 
\exp \left[ -\sum_{k=1}^{\infty} \frac{(-a\eps)^k}{k} S_k(n-1)\right] .
\eea
The product $\exp{(\eps\gamma_E)}\Gamma(1+\eps) = 1+\frac{1}{2}\zeta[2]\eps^2 + O(\eps^3)$ yields expressions with zeta numbers $\zeta[n]$, and, taking all terms together, one  gets a collection of inverse binomial sums\footnote{For the first four terms of the $\eps$-expansion in terms of inverse binomial sums or of polylogarithmic functions, see \cite{Gluza:2007bd}.}; 
the first of them is the IR divergent part: 
\bea
V(s) &=& \frac{V_{-1}(s)}{\eps} + V_0(s) + \cdots
\\
\label{vm1}
V_{-1}(s) &=& \frac{1}{2}\sum_{n=0}^{\infty}  \frac{s^n} { \binom{2n}{n} (2n+1)}
 =
 \frac{1}{2}\frac{4\arcsin(\sqrt{s}/2)}{\sqrt{4-s}\sqrt{s}}.
\eea

This procedure applies similarly to the Bhabha box diagram \cite{Fleischer:2006ht}. 
We take for definiteness the $s$ channel scalar loop integral.
The $F$-form is (again with $m=1$):
\bea
F (s,t) = [X[2] + X[4]]^2 + [- s]~X[1]X[3] + [-t] ~X[2]X[4],
\eea
and an MB representation is,
after continuation to small $\eps$, a sum of two terms:
\begin{eqnarray}
\label{integralsIR2} 
B(s,t) &=&
\frac{(-s)^{-\eps}}{2st}  
\frac{\Gamma[1+\eps] \Gamma[-\eps]^2}{\Gamma[-2\eps]} \frac{e^{\epsilon \gamma_E}}{2 \pi i }
    \int\limits_{-i \infty-7/16}^{+i \infty-7/16} d {z_1}
  (-t)^{-z_1}
\frac{{\Gamma}^3[-z_1] \Gamma[1+z_1]}
 { \Gamma[- 2 z_1]}
\\ \nonumber 
\label{twodimm}
&& +~
\frac{1}{t^2}
\frac{1}{\Gamma[-2\eps]} 
 \frac{e^{\epsilon \gamma_E}}{(2 \pi i )^2}
    \int\limits_{-i \infty-3/4}^{+i \infty-3/4} d {z_1}
 \left(\frac{s}{t}\right)^{z_1}
\Gamma[-z_1]\Gamma[-2(1+\eps+z_1)]
\Gamma[1+z_1]^2
\\\nonumber
    &&\times \int\limits_{ -i \infty-7/16}^{+i \infty-7/16} d {z_2}
(-t)^{-\eps-z_2}
\Gamma[-z_2]
\frac{ \Gamma[-1-\eps-z_1-z_2]^2}{\Gamma[-2(1+\eps+z_1+z_2)]}
  \Gamma[ 2 +\eps + z_1 + z_2].
\label{Int2}
\end{eqnarray}
Due to the pre-factors, 
$\Gamma[-\eps]^2/\Gamma[-2\eps] = -2/\eps + 2 \zeta[2] \eps + O(\eps^2)$ and $1 / \Gamma[-2\eps] = -2\eps+ 4 \gamma_E \eps^2 + O(\eps^3) $, 
only the first integral contributes to the first two terms of the $\eps$-expansion, 
\bea
B(s,t) &=& \left[ \frac{1}{\eps} - \ln(-s)\right] \frac{V_{-1}(t)}{(-s)} + O(\eps),
\eea
where $V_{-1}$ is from (\ref{vm1}),
and the IR divergency is:
\bea
B_{-1}(s,t) &=& \frac{V_{-1}(t)}{(-s)}.
\eea
We reproduce here the well-known fact that IR-divergences of vertices and boxes are algebraically related, see e.g. \cite{Gluza:2007uw}.

\section{Mixed virtual and real IR-singularities: massive Bhabha pentagon}
Things become more interesting for pentagon diagrams (se also \cite{Gluza:2007uw,Fleischer:2007ff,Fleischer:2007ph}).
We again use Bhabha scattering as an example. A compact $F$-form is:
 \bea
\label{fform}
F(s,t,t',V_2,V_4) &=&  (x_2+x_4+x_5)^2 + [-s] x_1 x_3 + [-{V_4}] x_3x_5 
+
 [-t]x_2x_4 + [-t']x_2x_5 +[-{V_2}]x_1x_4.
\nl
\eea
It exhibits a set of five invariants (out of a set of 10 scalar products at choice) describing the kinematics of a $2\to3$ process (here assuming a final state emission of an unresolved photon from an $s$ channel box diagram).
The $s,t,t'$ are the usual Mandelstam variables for the fermions in $e^+e^-\to e^+e^-\gamma$, and:
\bea
 V_i = 2 p_{f_i} p_{\gamma}, ~~~~i=1,\ldots 4.
\eea
The $V_i$ are proportional to the energy of the potentially unresolved massless particle.

{\em From a subsequent phase space integration, we have to expect endpoint singularities arising from terms proportional to $1/V_2\sim 1/E_{\gamma}$ and $1/V_4\sim 1/E_{\gamma}$, so we have to control, for a complete treatment of the IR-problem, not only the $\eps$-expansion, but also the first terms of the $ V_2,V_4$ expansions for small  $ V_2,V_4$.}

In fact, the $F$-form (\ref{fform}),  written here in its shortest form, depends on those two of the four $V_i$ which are related to the phase space of the given topology.

The MB-representation is a useful tool for that problem.
For the IR limit, we may approximate $t' = t$, and the scalar pentagon may be written as:
\bea
I_5 &=&
\frac{-e^{\eps\gamma_E}}{(2\pi i)^4}
\prod_{i=1}^4 \int\limits_{-i\infty+u_i}^{+i\infty+u_i} d z_i
 (-s)^{z_2} 
(-t)^{z_4} (-V_2)^{z_3} (-V_4)^{-3-\eps-z_1-z_2-z_3-z_4}
\frac{\prod\limits_{j=1}^{12}\Gamma_j}{\Gamma_0\Gamma_{13}\Gamma_{14}},
\nonumber\\
\eea
with $u_i=(-5/8,-7/8,...)$ and with a normalization 
$ \Gamma_{0}=\Gamma[-1-2\eps]$,
and the other $\Gamma$-functions are:
\bea\label{mbaux}
\Gamma_1&=&\Gamma[-z_1],
~~
\Gamma_2=\Gamma[-z_2],
~~
\Gamma_3=\Gamma[-z_3],
~~
\Gamma_4=\Gamma[1+z_3],
\nl
 \Gamma_5&=&\Gamma[1+z_2+z_3],
~~
\Gamma_6=\Gamma[-z_4],
~~
\Gamma_7=\Gamma[1 +z_4],
~~
 \Gamma_8=\Gamma[ -1-\eps - z_1 - z_2],
\nl
\Gamma_{9}&=&\Gamma[-2-\eps-z_1-z_2-z_3-z_4],
~~
\Gamma_{10}=\Gamma[-2-\eps-z_1-z_3-z_4],
\nl
 \Gamma_{11}&=&\Gamma[-\eps+z_1-z_2+z_4],
~
\Gamma_{12}=\Gamma[3 +\eps+z_1+z_2+z_3+z_4],
\eea
and, in the denominator:
\bea
\Gamma_{13}&=&\Gamma[-1-\eps-z_1-z_2-z_4] ,
~~
\Gamma_{14}=\Gamma[-\eps-z_1-z_2+z_4] .
\eea
Leaving out here the details of derivation (see \cite{Gluza:2007uw} for that), we just mention that we have to consider, after continuation in $\eps$, eleven MB-integrals, being at most 4-dimensional (for $t'=t$).  
The resulting IR-sensible part is:
\bea\label{i5ir}
I_5^{IR} &=& I_5^{IR}(V_2) +I_5^{IR}(V_4),
\\\label{i5ir2}
I_5^{IR} (V_i)&=& \frac{ I_{-1}^s(V_i)}{\eps} + I_0^s(V_i).
\eea
The $\eps$-pole is again proportional to that of the vertex:
\bea 
\label{sm1}
\frac{I_{-1}^s(V_i)}{\eps} &=&
\frac{1}{2sV_i\eps} \sum_{n=0}^{\infty} 
\frac{(t)^n}{ \begin{pmatrix}2n\\n\end{pmatrix} (2n+1)} 
~~=~~\frac{V_{-1}(t)}{sV_i\eps},
\eea
 and:
\bea\label{i0svi}
I_0^s(V_i) &=&\frac{1}{2sV_i}
 \sum_{n=0}^{\infty} 
\frac{(t)^n}{ \begin{pmatrix}2n\\n\end{pmatrix} (2n+1)}
\left[ -2\ln(-V_i) - 3 S_1(n) +2 S_1(2n+1)\right] ,
\eea
where we  have to understand $\ln(-V_i) = \ln(V_i/s) + \ln[-(s+i\delta)/m^2]$.
The series for $I_0^s(V_i)$  may be summed up in terms of polylogarithmic functions with the aid of Table~1 of Appendix D of \cite{Davydychev:2003mv}, see also \cite{Gluza:2007uw}.

{\em Equations (\ref{i5ir}) -- (\ref{i0svi}) are the main physical result of the study.
One may express the complete IR-divergent part of an amplitude with 5-point functions in terms of those expressions, subtract it from the complete, divergent amplitude and get a matrix element, which is integrable in four dimensions.}

In the rest of this short write-up 
we would like to demonstrate why  we have here besides the harmonic numbers $S_1(n)$ also those of the kind $S_1(2n+1)$. 
As mentioned, the scalar 5-point function may be written as a sum of eleven MB-integrals after $\eps$-continuation, before $\eps$-expansion.
Two types of them contribute to the IR-part (there are four such integrals [in an ad-hoc notations $J_3, J_4, J_7, J_9$], but with a symmetry $V_2 \leftrightarrow V_4$).
The first one is:
\bea
J_7&=&
- \frac{ (V_2/s)^{2\eps}}{s V_4} 
\Gamma[-2\eps]   \Gamma[1 + 2\eps]\frac{e^{\eps \gamma_E}}{2\pi i}
\int\limits_{-i\infty-5/8}^{+i\infty-5/8} dz (-t)^{-1 - z} \frac{ \Gamma[\eps - z]\Gamma[2\eps - z]
   \Gamma[-z]\Gamma[1 + z] }{\Gamma[2\eps - 2z]}
\nl
&=&
- \frac{(V_2/s)^{2\eps}} {s V_4} 
~~
\frac{ e^{\eps\gamma_E} \eps \sqrt{\pi} } {2^{2\eps}}
~~ \frac{\Gamma[-2\eps]\Gamma[2\eps] \Gamma[1 + 2\eps]    }{\Gamma[3/2 + \eps]}
~~
_2F_1[{1, 1 + 2\eps}, {3/2 + \eps}, t/4] 
\nl
&=&
-
\frac{ (V_2/s)^{2\eps}}{s V_4}  e^{\eps\gamma_E} 
~\Gamma[-2\eps]\Gamma[1 + 2\eps]
~\sum_{n=1}^{\infty}
t^{n-1}
\frac{ \Gamma[\eps + n]\Gamma[2\eps + n] }{
  \Gamma[2\eps + 2n]} .
\ea
The $J_7$ is proportional to $1/V_4$.
We have a second integral of the same type, being proportional to $1/V_2$:
\bea
J_3 &=& J_7(V_4\leftrightarrow V_2).
\eea
The other type of integrals $J_4,J_9$, with:
\bea
J_4 &=& J_9(V_4\leftrightarrow V_2),
\eea
 are two-fold MB-integrals:
\bea
\label{j9}
J_9 
&=&
\frac{\Gamma[-2 \eps]} { \Gamma[-1 - 2 \eps]}
\frac{e^{\eps\EulerGamma}} {(2\pi i)^2} 
\int\limits_{-i\infty-5/8}^{+i\infty-5/8}dz_1 \int\limits_{-i\infty-7/8}^{+i\infty-7/8}dz_2
(-s)^{z_2} (-t)^{\eps - z_1 + z_2}
   (-V_2)^{-1-2 \eps - z_2}  (-V_4)^{-2-z_2}
\nl
&&\times~
 \Gamma[-z_1]
   \Gamma_A[-1 - z_2] \Gamma[-2 \eps - z_2]
   \Gamma[-1 - \eps - z_1 - z_2]\Gamma_C[-\eps + z_1 - z_2]
   \Gamma[-z_2]
\nl&&\times~
 \frac{\Gamma[2 + z_2] ~~\Gamma_B[1 + 2 \eps + z_2]
   ~~\Gamma[1 + \eps - z_1 + z_2]
}{
 \Gamma[-2z_1]
   ~~\Gamma[-1 - 2 \eps - 2 z_2]} .
\eea
The integral looks like being, in the limit $V_4 \to 0$, too singular.
This limit is an endpoint of the phase space integration.
Let us close the contour to the left.
We {\em shift} now the integration contour in $z_2$ to the left, raising in this way the (real part of the) power of $(-V_4)$ to a value which makes the photon phase space integral explicitely integrable at $V_4 \to 0$ in $d=3-2\eps$ space dimensions.
If singularities of the integrand (from $\Gamma$-functions)  at some values $z_2=z_R$ are crossed one has to add the corresponding residues $J_9^{R}(z_R)$, so getting one-dimensional MB-integrals to be considered:
\bea
J_9 = 2 \pi i \sum_{z_R} J_9^{R}(z_R) + J_9^{shift}.
\eea
The resulting integral $J_9^{shift}$ differs from $J_9$ only by the shifted integration path, but will now not contribute to the IR-singular part and will not be considered here any more.
We see that only the residues of crossed singular points in $z_2$ contain the IR-relevant endpoint singularities in $V_2, V_4$.
Here, two of them (at $z_2= -1$ [argument of $\Gamma_A$ in (\ref{j9})] and at 
$z_2= -1 - 2 \eps$ [argument of $\Gamma_B$in (\ref{j9})]) contribute due to a shift from $\Re z_2=-7/8 $ to $\Re z_2 =-7/8-1=-15/8$.
The first of them is: 
\bea\label{res09A}
J_{9A}
&=&
-  \frac{(-V_2)^{-2 \eps}} {sV_4}
\frac{\Gamma[-2 \eps] \Gamma_B[2 \eps]}{\Gamma[-1 - 2 \eps]}
\frac{e^{\eps\EulerGamma}} {2\pi i} 
\nl
&&\times~
\int\limits_{-i\infty-5/8}^{+i\infty-5/8} dz_1 (-t)^{-1-z_1}
\frac{  \Gamma[-2 \eps - z_1] \Gamma[-\eps - z_1] \Gamma[-z_1]
   \Gamma_C[1 + z_1])}{   \Gamma[-1 - 2 \eps] \Gamma[-2\eps -2 z_1)]}
\nl
&=&
\label{resid09A}
\frac{(-V_2)^{-2 \eps} 2^{2 \eps}} {sV_4} e^{\eps\EulerGamma} 
\frac{\eps \sqrt{\pi}   \Gamma[-2 \eps]^2 \Gamma_B[2 \eps]} {\Gamma[-1 - 2 \eps]
   \Gamma[3/2 - \eps]}
~~~ _2F_1[{1, 1 - 2 \eps}, {3/2 - \eps}, t/4] .
\eea
We performed here an irrelevant shift $z_1 \to z_1 + \eps$ in order to make the argument of $\Gamma_C$ independent of $\eps$. 
This will be here the only $\Gamma$-function producing residues in $z_1$ when closing again the contour to the left.
The explicit $\eps$-expansion of the hypergeometric function in (\ref{resid09A}) may be obtained with the Mathematica package HypExp2 \cite{Huber:2005yg,Huber:2007dx}.
Alternatively, we may perform the sum of residues arising from $\Gamma_C(1+z_1)$ directly:
\bea\label{resid09aux0A}
J_{9A}
&=&
- \frac{(-V_2)^{-2 \eps}} {sV_4}
e^{\eps\EulerGamma} 
\frac{\Gamma[-2 \eps]  \Gamma[2 \eps]}{\Gamma[-1 - 2 \eps]}
\sum_{n=1}^{\infty}
t^{n-1}  \frac{ \Gamma[-2 \eps + n]
    \Gamma[-\eps + n]}{    \Gamma[-2\eps +2 n]},
\eea
and apply then (\ref{harmsum}) to it.
It is here where we may see why the harmonic numbers $S_1(2n+1)$ appear, which were not contributing to the vertex or the box:
There was no $\Gamma$-function with an $\eps$-shifted,  doubled argument in the denominator of the final sums, while here this appears.
In the general massive case, this will usually happen.  
 
The second residue crossed by the contour shifting in the $z_2$-plane gives the third kind of contribution to be added; after a shift $z_1 \to z_1 - \eps$: 
\bea\label{res09B}
J_{9B} &=&
\frac{(V_4/s)^{2 \eps}} {sV_4}
    \Gamma[1 - 2 \eps] \Gamma[-2 \eps] \Gamma_A[2 \eps]
\frac{e^{\eps\EulerGamma}} {2\pi i}
\int\limits_{-i\infty-5/8}^{+i\infty-5/8} dz_1
  (-t)^{-1-z_1}
\nl && \times~
\frac{  \Gamma[\eps - z_1] \Gamma[2 \eps - z_1] \Gamma[-z_1]
    \Gamma_C[1 + z_1]
}{
    \Gamma[-1 - 2 \eps] \Gamma[2\eps -2 z_1]}
\nl
&=&
\label{resid09B}
\frac{(V_4/s)^{2 \eps}} {sV_4}
e^{\eps\EulerGamma} 
\eps\sqrt{\pi}
\frac{   \Gamma[1 - 2 \eps] \Gamma[-2 \eps]\Gamma[2 \eps]^2
} {2^{2 \eps} \Gamma[-1 - 2 \eps] \Gamma[3/2 + \eps]}
~~~_2F_1 [{1, 1 + 2 \eps}, {3/2 + \eps}, t/4]
\nl\label{res09aux0Bout}
&=&
\frac{(V_4/s)^{2 \eps}} {sV_4}
e^{\eps\EulerGamma}
\frac{     \Gamma[1 - 2 \eps] \Gamma[-2 \eps]\Gamma[2 \eps]}
     {     \Gamma[-1 - 2 \eps]}
\sum_{n=1}^{\infty}t^{n-1}
\frac{ \Gamma[\eps + n]   \Gamma[2 \eps + n]}{\Gamma[2\eps +2 n]}
\eea
Again, this may be expanded into an $\eps$-series over inverse binomial sums by use of (\ref{harmsum}).

Collecting everything together, we rediscover (\ref{i5ir}) (plus additional terms of no relevance for the IR-treatment):
\bea
J_3 + J_{4A} + J_{4B} + J_7 + J_{9A}+J_{9B} 
&=& 
\frac{1}{\eps}\left[  I_{-1}^s(V_2) + I_{-1}^s(V_4)\right] + I_0^s(V_2) + I_0^s(V_4) +\cdots
\eea

\section{Conclusions}
We gave a pedagogical introduction to the treatment of mixed real and virtual IR-singularities.
This kind of problems arises in NNLO problems, where one has to treat the unresolved massless particle phase space for loop integrals.
The presented method was exemplified for a scalar integral, but it may be easily applied to general tensor functions.
For the QED pentagon, this is discussed in \cite{Gluza:2007uw}, which may be considered as an introduction to this presentation.
A {\em derivation} of MB-representations for higher $n$ point functions or for multi-loop integrals is more or less straightforward, although an {\em analytical evaluation} will become more and more troublesome.
It is an interesting open question how useful the MB techniques might appear for realistic, so far unsolved applications.

\acknowledgments{We would like to thank J. Fleischer for useful discussions.
\\
The present work is supported in part 
by the European Community's Marie-Curie Research Training Networks  MRTN-CT-2006-035505 `HEPTOOLS'
%
and MRTN-CT-2006-035482 `FLAVIAnet', 
and by  
Sonderforschungsbe\-reich/Trans\-regio 9--03 of Deutsche Forschungsgemeinschaft
`Computergest{\"u}tzte Theo\-re\-ti\-sche Teil\-chen\-phy\-sik'. }

\providecommand{\href}[2]{#2}\begingroup\raggedright\endgroup

\end{document}